# Learning hidden elasticity with deep neural networks

Chun-Teh Chen[1] and Grace X. Gu[2]*

[1] Department of Materials Science and Engineering, University of California, Berkeley, CA 94720, USA

[2] Department of Mechanical Engineering, University of California, Berkeley, CA 94720, USA

*Address correspondence to: ggu@berkeley.edu, +1.510.643.4996

**Abstract**

We introduce a *de novo* elastography method to learn the elasticity of solids from measured strains. The deep neural network in our new method is supervised by the theory of elasticity and does not require labeled data for training. Results show that the proposed method can learn the hidden elasticity of solids accurately and is robust when it comes to noisy and missing measurements. A probable elasticity distribution for areas without measurements may also be reconstructed by the neural network based on the elasticity distribution in nearby regions. The neural network learns the hidden elasticity of solids as a function of positions and thus it can generate elasticity images with an arbitrary resolution. This feature is applied to create super-resolution elasticity images in this study. We demonstrate that the neural network can also learn the hidden physics when strain and elasticity distributions are both given. The proposed method has various unique features and can be applied to a broad range of elastography applications.

**Introduction**

Being able to image the material property distribution of solids non-invasively has a broad range of applications in materials science, biomechanical engineering, and clinical diagnosis. For instance, as various disease progresses, the elasticity (quantified as the Young's modulus or shear modulus) of human cells, tissues, and organs may be altered significantly[1]. Palpation (e.g., breast self-examination) utilizes the difference between the elasticity of healthy and cancerous tissues to distinguish them. Elasticity imaging, known as elastography, is an emerging method to qualitatively image the elasticity distribution of an inhomogeneous body[2-5]. A long-standing goal of elastography is to provide alternative methods of clinical palpation for reliable tumor diagnosis. The displacement distribution of a body under externally applied forces (or displacements) can be acquired by a variety of imaging techniques such as ultrasound, magnetic resonance, and digital image correlation[6,7]. A strain distribution, determined by the gradient of a displacement distribution, can be computed (or approximated) from measured displacements. If the strain and stress distributions of a body are both known, the elasticity distribution can be computed using the constitutive elasticity equations (Hooke's law). However, there is currently no technique that can measure the stress distribution of a body *in vivo*. Therefore, in elastography, the stress distribution of a body is commonly assumed to be uniform and a measured strain distribution can be interpreted as a relative elasticity distribution. This approach is referred to as strain-based elastography and has the advantage of being easy to implement. The uniform stress assumption in this approach, however, is inaccurate for an inhomogeneous body. The stress field of a body can be distorted significantly near a hole, inclusion, or wherever the elasticity varies. This phenomenon, known as stress concentration, is of great interest in industry and academia. Though the strain-based elastography has been deployed on many commercial ultrasound diagnostic-imaging devices, the elasticity distribution predicted based on this method is prone to inaccuracies[8]. To mitigate this misinterpretation, a research field focusing on solving an inverse problem associated with elasticity imaging has been extensively investigated for decades. In this approach, referred to as model-based elastography, the elasticity distribution of a body may be, in principle, recovered by modeling its elastic behavior and solving an inverse problem.

In this study, an inverse problem in elasticity is formulated: to determine the elasticity distribution of a body from a given displacement (or strain) distribution. It is an inverse problem since a typical forward problem in elasticity is to determine the displacements of a body from a given elasticity distribution. To tackle this inverse problem, the dominant approaches in the literature are based on minimizing the difference between measured and simulated displacements[8-11]. However, these approaches are computationally expensive as they require many iterations and each iteration requires solving a forward problem and conducting sensitivity analysis using the finite element method (FEM). It is possible to solve this inverse problem directly. In other approaches, known as direct approaches, measurements are considered as coefficients in the partial differential equations (PDEs) of equilibrium[12-14]. Though these direct approaches are computationally efficient, they may perform poorly when measurements contain large strain gradients or noise, or the elasticity distribution is not smooth. Moreover, the error from noise tends to propagate along the integration path when solving the PDEs and thus causing inaccurate predictions. Due to these limitations, most model-based elastography methods were only applied to solve simple problems in elasticity imaging such as a uniform soft body containing a few hard inclusions.

Inverse problems arise in many scientific and engineering fields and are typically difficult to solve by conventional approaches[15-18]. Much progress towards artificial intelligence (AI) and machine learning (ML) have been made and provided novel directions to solve these inverse problems[19-22]. For instance, ML techniques were applied to solve inverse problems in materials design[23-25], fluid mechanics[26], and PDEs[27]. In this study, we consider the possibility of applying ML techniques to solve the inverse problem in elasticity. To obtain useful information from an elasticity image, the number of pixels (resolution) is typically on the order of $10^3$ to $10^5$. Due to the high-dimensional input and output spaces, the hidden correlation between the strain and elasticity distributions of a body is difficult to be captured by supervised learning using labeled data. In principle, supervised learning may work for this inverse problem if the number of possible elasticity distributions can be reduced. A simple way to do so is, for instance, to consider a uniform soft body containing a few hard inclusions and to constrain the shapes, sizes, locations, and elastic moduli of these hard inclusions. However, adding such artificial constraints to the inverse problem may not make sense in practice.

In this study, we introduce a *de novo* elastography method to learn the hidden elasticity of a body with a deep neural network. Our new method is not supervised by labeled data; it is supervised by the theory of elasticity. The constitutive elasticity equations and equilibrium equations for solving the inverse problem are encoded in the neural network. Our new method is a general-purpose approach for elastography. No artificial constraint on possible elasticity distributions is imposed. We show that the proposed method can accurately reconstruct the elasticity distribution of a body from a given strain distribution. The proposed method is robust when it comes to noisy and missing measurements. Moreover, the proposed method can predict a probable elasticity distribution for areas without data based on the elasticity distribution in nearby regions. This ability can be used to generate super-resolution elasticity images. We demonstrate that the proposed method not only can learn the hidden elasticity but also can decrypt the hidden physics of the inverse problem when strain and elasticity distributions are both given. The unique features of the proposed method make it a promising tool for a broad range of elastography applications.

**Results**

**Deep learning for elastography:** Our new method uses a deep neural network to learn the hidden elasticity of a body from a given strain distribution. The flowchart of the proposed method is shown in **Fig. 1**. Since biological tissues are almost incompressible[7], each material point on the region of interest is assumed to be linear, isotropic, and incompressible. Thus, the elasticity at a material point can be described by only one elastic constant, either the Young's modulus $E$ or shear modulus $G$. Without loss of generality, we choose the Young's modulus $E$ to quantify the elasticity. We consider the region of interest in a plane stress state. The relevant components of the stress are $\sigma_{xx}$, $\sigma_{yy}$, and $\tau_{xy}$. A measured strain distribution includes three strain images ($\varepsilon_{xx}, \varepsilon_{yy}, \gamma_{xy}$). The stain information is converted to a dataset, in which each data point (corresponding to each material point) contains the position and strain of the point. The neural network takes the position and strain at a material point and predicts the elasticity $E$ of the point. For the proposed method, the derivatives of strains are not required for solving the inverse problem.

Thus, it is unnecessary (though it is doable) to train the neural network to predict the strain. The stress at a material point is calculated by the encoded elastic constitutive relation from the measured strain and predicted elasticity jointly (see Methods). After the entire dataset is passed forward through the neural network, a predicted stress distribution is generated.

A predicted stress distribution includes three stress images ($\sigma_{xx}, \sigma_{yy}, \tau_{xy}$). Before training, these strain images are unlikely to satisfy the conservation of linear momentum (equilibrium) as the initial elasticity distribution $E(x, y)$ is generated by random initialization of weights and biases in the neural network. To evaluate how close the predicted stress distribution is from equilibrium, the stress images are passed forward through a convolutional layer. Unlike other convolutional neural networks, in which filter values (or weights) need to be learned from labeled data, the filter values in our convolutional layer are encoded in such a way that the convolution operation can be used to evaluate the conservation of linear momentum (to be discussed in the next section). Residual (unbalanced) force maps are generated after the convolution operation. The training process minimizes the residual forces and updates the prediction on the elasticity distribution using backpropagation.

**Learning the conservation of linear momentum:** We consider a small cube with sides of length $h$. The conservation of linear momentum for the cube can be expressed as (see Methods):

$$\sigma_{xx}(x + h, y) - \sigma_{xx}(x, y) + \tau_{yx}(x, y + h) - \tau_{yx}(x, y) = 0 \tag{1}$$

$$\sigma_{yy}(x, y + h) - \sigma_{yy}(x, y) + \tau_{xy}(x + h, y) - \tau_{xy}(x, y) = 0 \tag{2}$$

Let the cube contain 3-by-3 material points as shown in **Fig. 2a**. The conservation of linear momentum for the cube can be expressed in terms of the stresses at the material points:

$$\sum_{a=1}^{3}\sum_{b=1}^{3} w_{xx}(a,b)\sigma_{xx}(a,b) + w_{yy}(a,b)\sigma_{yy}(a,b) + w_{xy}(a,b)\tau_{xy}(a,b) = 0 \quad (3)$$

where $w_{xx}$, $w_{yy}$, and $w_{xy}$ are the convolution kernels for $\sigma_{xx}$, $\sigma_{yy}$, and $\tau_{xy}$, respectively. By choosing proper sets of values for the kernels, (3) can be used to describe the conservation of linear momentum for the cube. These sets of values are then encoded in the kernels to describe the equilibrium conditions in the *x*-direction and *y*-direction, respectively (**Fig. 2a**). These kernels are used as "equilibrium" detectors in the neural network and the convolution operation generates residual force maps, in which each element is calculated by:

$$e(i,j) = \sum_{a=1}^{3}\sum_{b=1}^{3} w_{xx}(a,b)\sigma_{xx}(i+a-1,j+b-1) + w_{yy}(a,b)\sigma_{yy}(i+a-1,j+b-1) \quad (4)$$
$$+ w_{xy}(a,b)\tau_{xy}(i+a-1,j+b-1)$$

The conservation of linear momentum is encoded in the neural network as domain knowledge to solve the inverse problem. Here, we consider the possibility of learning this domain knowledge from labeled data. To test this idea, an inhomogeneous body is modeled by the finite element method (FEM) with a 128-by-128 mesh. The elasticity field of the body is set to be a two-dimension sinusoidal function, given by:

$$E(x,y) = 0.45\left(1 + \cos\left(\frac{4\pi x}{L}\right)\cos\left(\frac{4\pi y}{L}\right)\right) + 0.1, \quad x \in [0,L], \quad y \in [0,L] \quad (5)$$

where *L* is the length of the body in both *x*-direction and *y*-direction. The unit is set to be megapascal (MPa). Consequently, the maximum and minimum Young's moduli are 1 MPa and 0.1 MPa, respectively. Since the body is assumed to be linear elastic, the absolute values of the Young's moduli are not important to the inverse problem. This "sinusoidal" model is subjected to externally applied displacements on the boundary. An average normal strain of 1% in the *x*-direction is caused by the displacement boundary condition. Similar finite element analysis was adopted in our previous works[23]. The elasticity and strain distributions of the body are shown in **Fig. 2b**. The elasticity and strain distributions are both fed into the neural network and the loss function is set to be the mean absolute error (MAE) in the residual forces:

$$\text{loss} = \frac{1}{m^2} \sum_i^m \sum_j^m |e(i,j)| \quad (6)$$

The objective is to learn the convolution kernels to describe the conservation of linear momentum. From (4) and (6), it can be seen that the kernels cannot be uniquely determined by minimizing the residual forces as there is an infinite number of kernel combinations that can produce zero residual forces. A trivial solution is to set all kernel values to zeros. Thus, to obtain a physically meaningful result, additional information must be given. For instance, when the kernel for $\tau_{xy}$ to describe the conservation of linear momentum in the *x*-direction is given, the other two kernels, which are for $\sigma_{xx}$ and $\sigma_{yy}$, can be learned. Similarly, when the kernel for $\tau_{xy}$ to describe the conservation of linear momentum in the *y*-direction is given, the other two kernels can be learned. The kernels learned by the neural network are shown in **Fig. 2b** and these kernels are almost identical to those derived mathematically. The result shows that the kernels encoded in the neural network to describe the conservation of linear momentum are accurate as the same kernels can be learned from the hidden correlation between the elasticity and strain distributions.

**Effect of physical constraints:** The inverse problem of elasticity formulated so far is ill-posed. Without additional information, no unique solution (elasticity distribution) can be generated from a given strain distribution. A trivial solution is an elasticity distribution being zero everywhere. To obtain a unique solution, one or more physical constraints must be imposed. A possible constraint can be, for instance, a specified value of the elasticity at a material point, the traction on the boundary, or the mean elasticity[13,14].

An absolute elasticity distribution may be reconstructed by solving the inverse problem when the specified value used in the constraint is a real physical quantity of the body; a relative elasticity distribution may be reconstructed when the specified value is not a real physical quantity but an arbitrary value. In this section, we investigate the effect of imposing different physical constraints on prediction accuracy. We consider a uniform soft body containing a circular hard inclusion. This "inclusion" model has been widely investigated in the literature to evaluate the performance of elastography methods. Here, the Young's moduli of the hard inclusion and soft body are set to be 1 MPa and 0.1 MPa, respectively.

The elasticity and strain distributions of the inclusion model are shown in **Fig. 3a**. The first scenario is assuming that the traction on the boundary is known. This information, however, may be difficult to obtain in practice. The second scenario is assuming that the mean elasticity is known. This scenario may be more reasonable in practice as the mean elasticity can be estimated by making some approximations on the elasticity distribution. Even if the mean elasticity is unavailable, one may specify it to an arbitrary value and obtain a relative elasticity distribution. We consider these two scenarios by adding the corresponding physical constraints into the loss function, respectively. Our new method takes a strain distribution and outputs predicted elasticity distributions based on either knowing the traction on the boundary or mean elasticity. The predicted elasticity distributions and relative error maps (as compared with the true elasticity distribution) are shown in **Fig. 3b** and the convergence of error over training is shown in **Fig. 3c**. In the first scenario, the traction on the boundary is known. Three force boundary conditions (BCs) can be applied to constrain the distribution of internal stresses as shown in the inner figure of **Fig. 3c**. The predicted elasticity distribution is very accurate with a mean relative error of 0.24%, while larger errors occur on the boundary between the soft body and hard inclusion due to a large elasticity difference. In the second scenario, the prediction is also very accurate with a mean relative error of 0.78%. As with the first scenario, larger errors occur on the boundary between the soft body and hard inclusion. The results suggest that applying a physical constraint based on the traction on the boundary may produce more accurate predictions. However, the traction on the boundary is in general difficult to obtain in practice. Thus, we only apply a physical constraint based on the mean elasticity to solve the following elasticity imaging problems.

**Effect of noise in measurements:** Measured strains naturally contain a certain amount of noise. In this section, we investigate the effect of noise in measurements on prediction accuracy. The "sinusoidal" model that is used to learn the conservation of linear momentum (**Fig. 2b**) is considered again here. The elasticity and strain distributions of the model are shown in **Fig. 4a**. In the first scenario, it is assumed that there is no noise in the measured strains. In the second scenario, white Gaussian noise is added to the measured strains. The level of noise is set to be approximately 3% of the measurements, which corresponds to a signal to noise ratio (SNR) of 30 dB. The predicted elasticity distributions and relative error maps are shown in **Fig. 4b** and the convergence of error over training is shown in **Fig. 4c**. In the first scenario, the predicated elasticity distribution is very accurate with a mean relative error of 0.41%. The error is uniformly distributed as the elasticity distribution is smooth. In the second scenario, the predicated elasticity distribution is still accurate with a mean relative error of 3.55%, given that the measured strains contain 3% white Gaussian noise (inner figure of **Fig. 4c**). The noise does not cause a catastrophic failure in learning the hidden elasticity. The results show that our new method is robust when it comes to noisy measurements.

**Effect of missing data in measurements:** In this section, we investigate the effect of missing data in measurements on prediction accuracy. Noisy measurements may still provide useful information to learning the hidden elasticity as discussed in the previous section. However, missing measurements (in some areas) not only provide no information to solve the inverse problem but also may make the inverse problem unstable. We consider a uniform hard body containing a soft inclusion with the shape of the UC Berkeley Cal logo. In this "Cal" model, the Young's moduli of the hard inclusion and soft body are set to be 1 MPa and 0.1 MPa, respectively. The elasticity and strain distributions of the model are shown in **Fig. 5a**. In the first scenario, it is assumed that there is no missing data in the measured strains. In the second scenario, the measured strains in an arbitrary area are removed (set to be zeros). The predicted elasticity distributions and relative error maps are shown in **Fig. 5b** and the convergence of error over training is shown in **Fig. 5c**.

In the first scenario, the predicated elasticity distribution is accurate with a mean relative error of 4.01%. This error is higher than that observed in the inclusion model (0.78%) due to the more complex inclusion shape in the Cal model. As with the inclusion model, larger errors occur on the boundary between the hard body and soft inclusion due to a large elasticity difference. In the second scenario, the predicated elasticity distribution is still accurate with a mean relative error of 7.14%, given that the measurements on a squared area (corresponds to 6.25% of the total area) are missing (yellow-boxed area in **Fig. 5b**). Larger errors occur on the boundary between the hard body and soft inclusion as well as on the area without data. If we exclude the area without data, the mean relative error is reduced to only 4.08%, almost the same as the error observed in the first scenario with the full data (4.01%). The results show that, for our new method, missing measurements only reduce the prediction accuracy in areas without data and do not affect the prediction accuracy in the remaining areas. For other elastography methods, no prediction can be made for areas without data. Our new method can predict a probable elasticity distribution for areas without data based on the elasticity distribution on nearby areas (inner figure of **Fig. 5c**).

**Super-resolution elasticity imaging:** For other elastography methods, the resolution of predicted elasticity distributions (elasticity images) depends on the resolution of measurements. Our new method, in theory, can generate elasticity images with any resolution. The neural network is not trained to learn the elasticity image of a body with a fixed resolution but to learn the elasticity field as a function of positions (**Fig. 1**). In this section, we investigate the possibility of generating super-resolution elasticity images from low-resolution measurements. We consider an inhomogeneous body with an elasticity distribution based on the *Mona Lisa* by Leonardo da Vinci. In this "Mona Lisa" model, the maximum and minimum Young's moduli are set to be 1 MPa and 0.1 MPa, respectively. The elasticity and strain distributions of the model are shown in **Fig. 6a**. The predicted elasticity distributions and relative error maps are shown in **Fig. 6b** and the convergence of error over training is shown in **Fig. 6c**. While the Mona Lisa model has an extremely complex elasticity distribution, the prediction accuracy is very high with a mean relative error of 2.49%. No difference between the ground truth and prediction can be seen with the naked eye. To understand how the neural network learns the hidden elasticity, intermediate predictions are shown in the inner figures of **Fig. 6c**. Interestingly, it can be seen that the neural network draws an outline first and then adds more details over training. This process is similar to how an artist draws. The Mona Lisa model is discretized with a 128-by-128 mesh. Thus, the resolution of the measured strains is also 128-by-128 (**Fig. 6a**). After learning the hidden elasticity from the measured strains, the neural network can generate elasticity images with any resolution. Here, the neural network is used to generate an elasticity image with a resolution of 512-by-512. For comparison, a crop of the ground truth image and that of the super-resolution image are shown in **Fig. 6d**. It can be seen that the super-resolution image is realistic and provides much more details compared with the ground truth image.

**Discussion**

We demonstrate that our new method can accurately learn the hidden elasticity of a body from a given strain distribution. The prediction accuracy depends on the complexity of the hidden elasticity. A higher prediction accuracy may be expected when the elasticity distribution is simpler or smoother. To make an accurate prediction on the hidden elasticity of the Mona Lisa model (**Fig. 6b**), the proposed method takes about 80 minutes for running 800,000 epochs on a single NVIDIA Tesla V100 GPU. However, after about 3,000 epochs (20 seconds) of training, an intermediate prediction with enough details can already be obtained (inner figures of **Fig. 6c**). With further improvements, the proposed method may have the potential to be used in an environment requiring real-time elasticity imaging.

**Figures:**

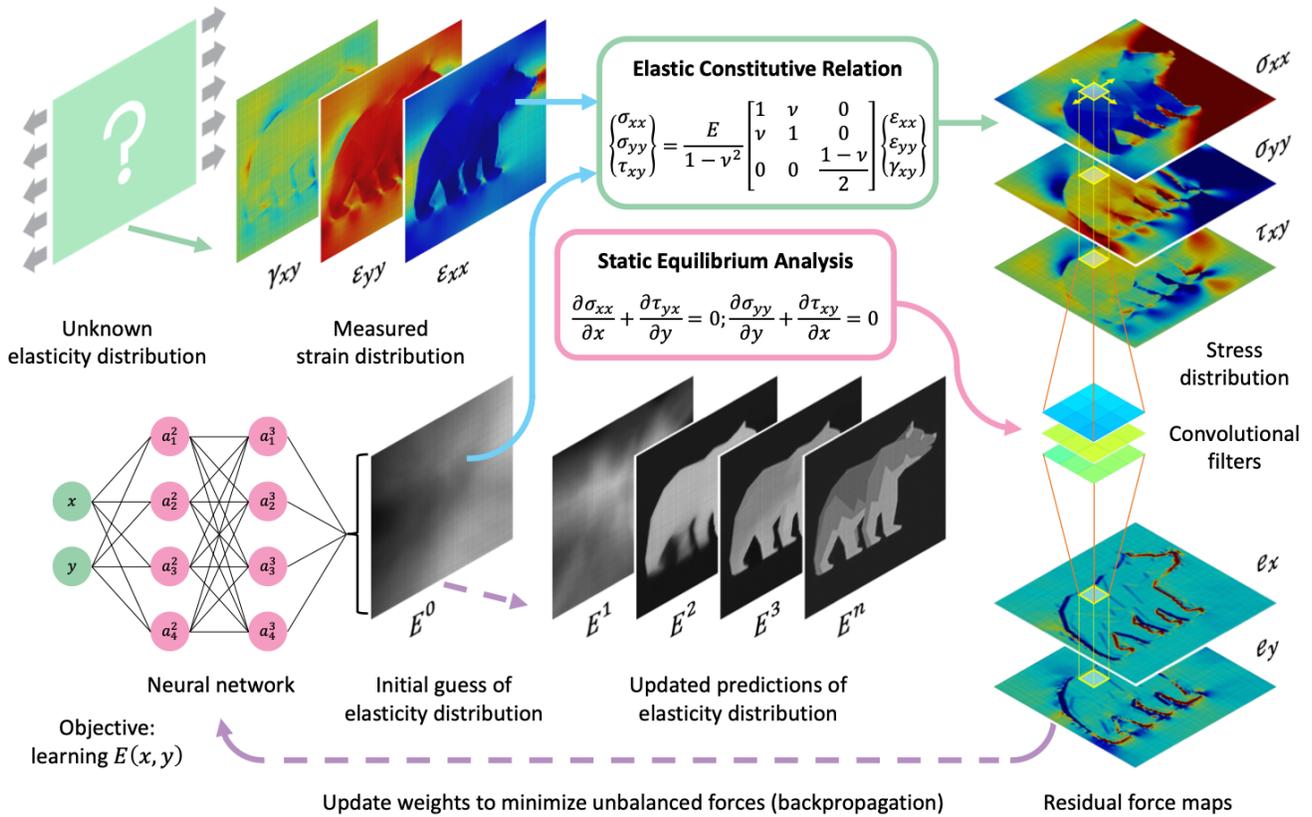

**Fig. 1| Flowchart of our new method for elastography.** The neural network takes the position and strain at a material point and predicts the elasticity of the point. The stress at a material point is calculated by the encoded elastic constitutive relation from the measured strain and predicted elasticity jointly. After the entire dataset is passed forward through the neural network, a predicted stress distribution is generated. The stress images are passed forward through a convolutional layer. Residual force maps are generated after the convolution operation. The training process minimizes the residual forces and updates the prediction on the elasticity distribution using backpropagation.

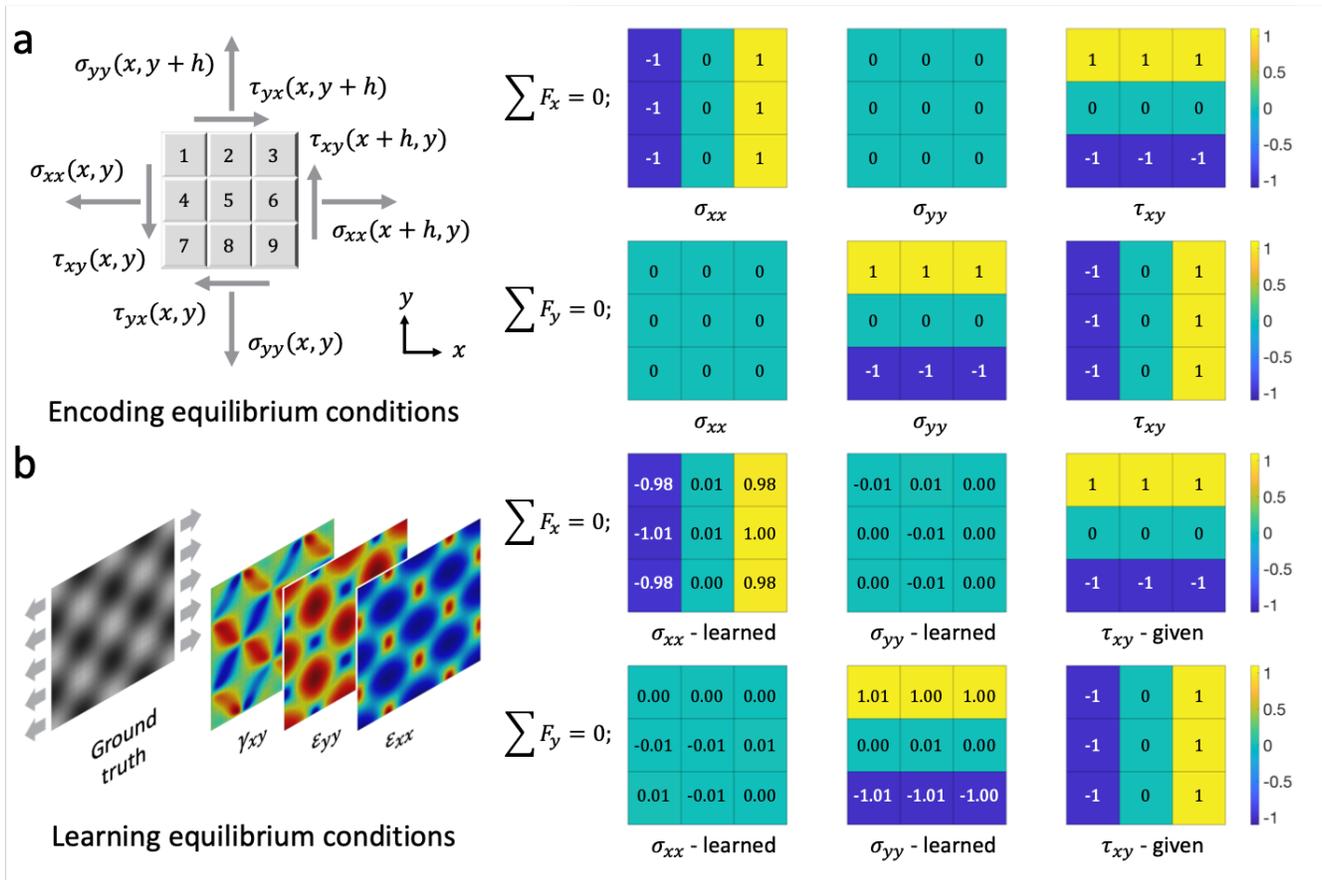

**Fig. 2| Encoding and learning equilibrium conditions.** (a) A representative cube consists of 3-by-3 material points. The conservation of linear momentum for the cube can be expressed in terms of the stresses at the material points. Two sets of kernels are encoded to describe the equilibrium conditions in the *x*-direction (upper three kernels) and *y*-direction (lower three kernels), respectively. (b) The sinusoidal model is subjected to externally applied displacements on the boundary. The elasticity and strain distributions are both fed into the neural network to learn the hidden equilibrium conditions. The kernels learned by the neural network are almost identical to those derived mathematically in (a).

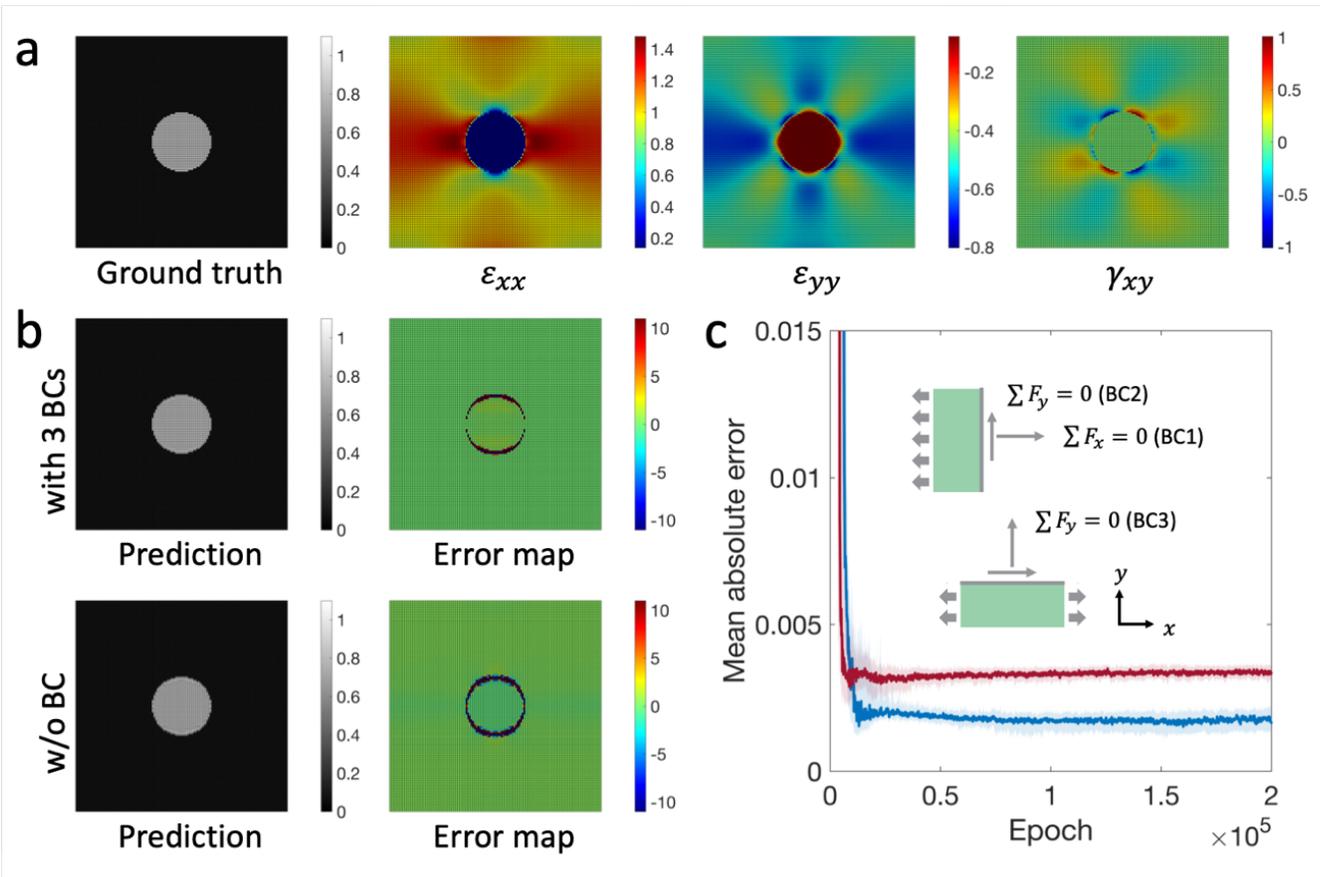

**Fig. 3| Effect of physical constraints.** (a) The elasticity and strain distributions of the inclusion model. The unit in the elasticity image is MPa and that in the strain images is the percentage (%). (b) The predicted elasticity distributions and relative error maps for the scenarios when the traction on the boundary is known (with 3 BCs) and when the mean elasticity is known (without BC), respectively. The unit in the error maps is the percentage (%). (c) The blue and red shaded lines represent the mean absolute error over training for the scenario with 3 BCs and that without BC, respectively. The results are obtained after training the neural network 100 times with different initial weights and biases. The line and shading represent the median and inter-quartile range, respectively. The inner figure shows the 3 BCs applied to constrain the distribution of internal stresses.

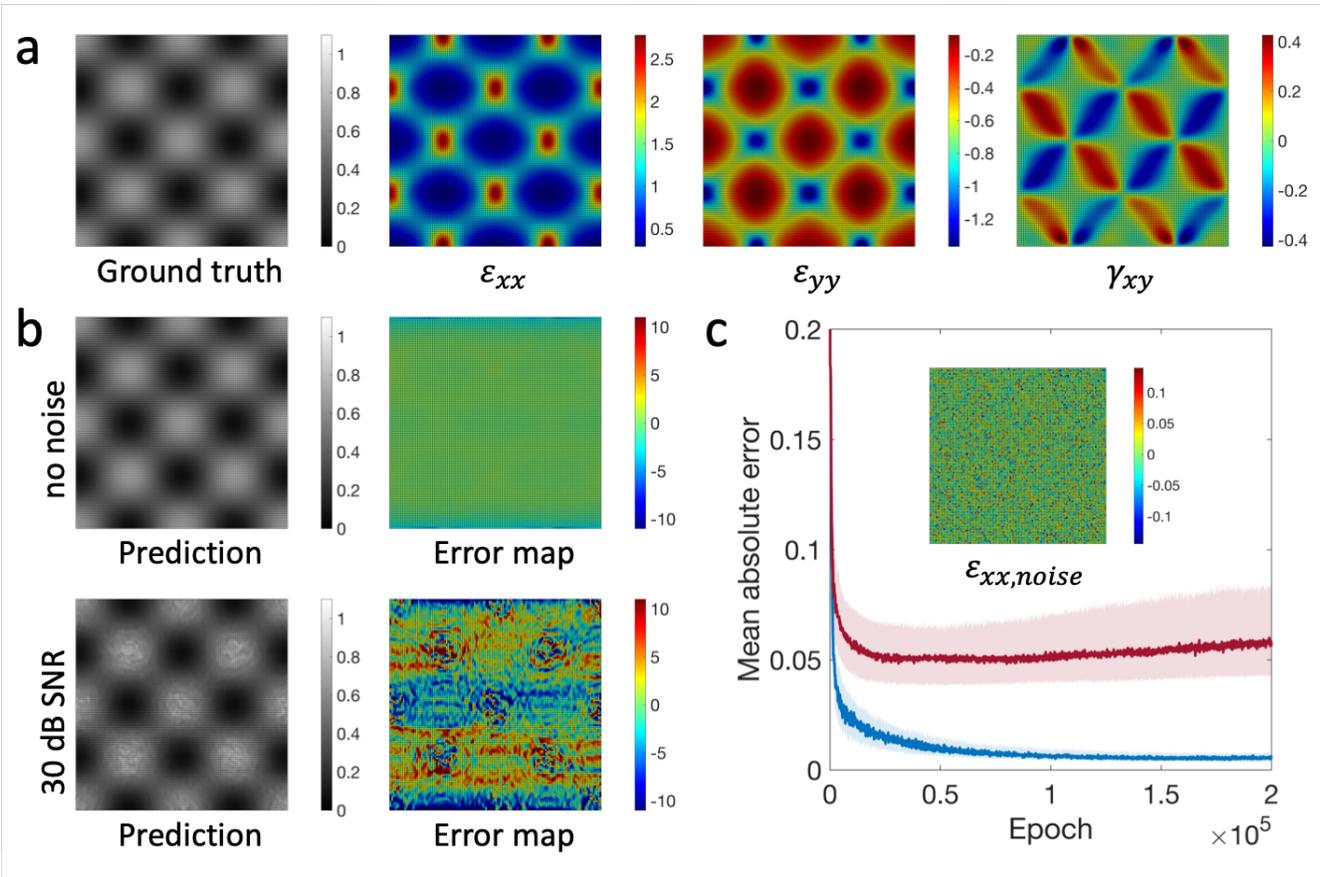

**Fig. 4| Effect of noise in measurements.** (a) The elasticity and strain distributions of the sinusoidal model. The unit in the elasticity image is MPa and that in the strain images is the percentage (%). (b) The predicted elasticity distributions and relative error maps for the scenarios when there is no noise in the measured strains and when white Gaussian noise of 30 dB SNR is added to the measured strains, respectively. The unit in the error maps is the percentage (%). (c) The blue and red shaded lines represent the mean absolute error over training for the scenario with no noise and that with noise, respectively. The results are obtained after training the neural network 100 times with different initial weights and biases. The line and shading represent the median and inter-quartile range, respectively. The inner figure shows the noise added to the strain of $\varepsilon_{xx}$. Due to the space, the noise added to the strains of $\varepsilon_{yy}$ and $\gamma_{xy}$ is not shown.

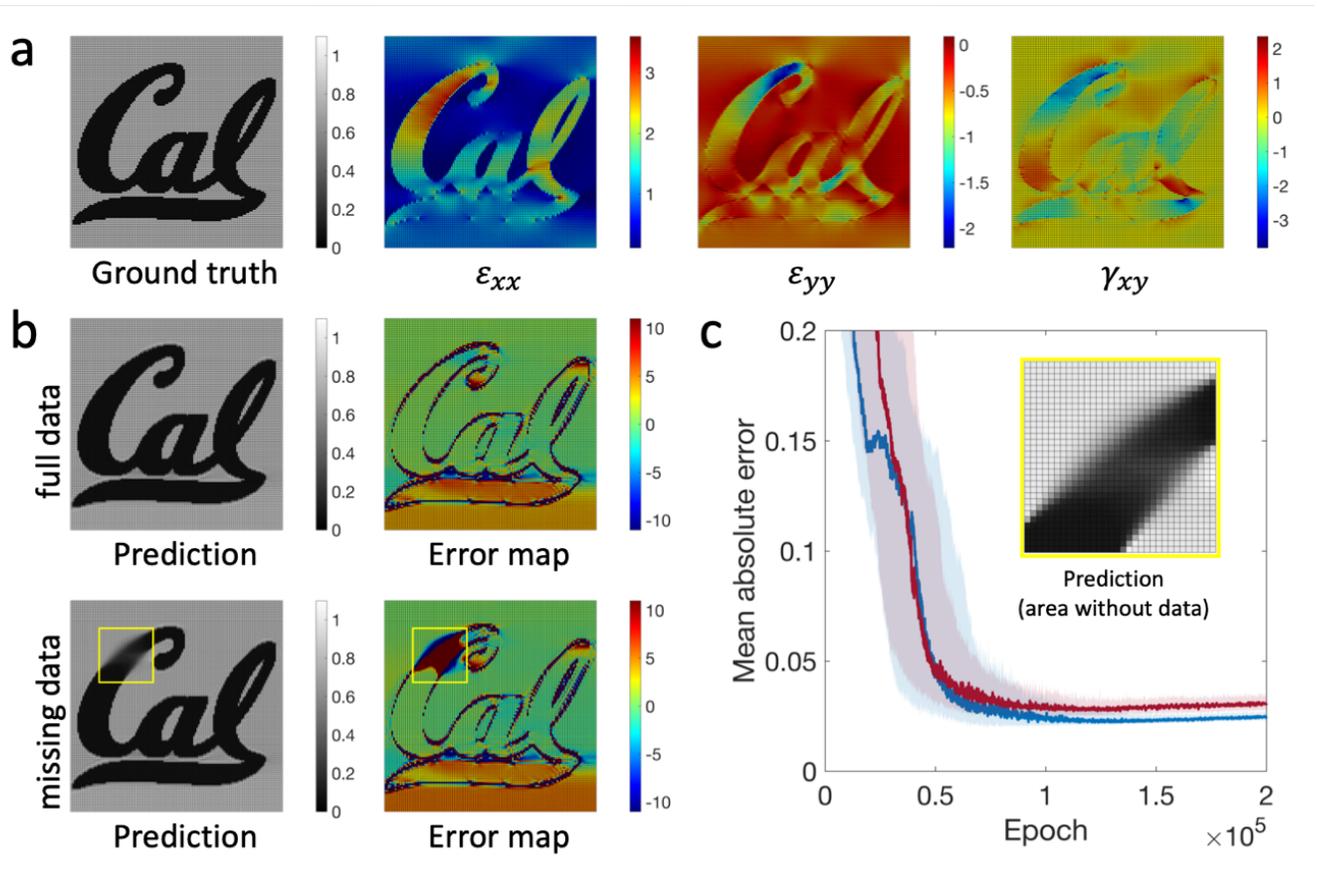

**Fig. 5| Effect of missing data in measurements.** (a) The elasticity and strain distributions of the Cal model. The unit in the elasticity image is MPa and that in the strain images is the percentage (%). (b) The predicted elasticity distributions and relative error maps for the scenarios when the full data is available and when there is missing data, respectively. The unit in the error maps is the percentage (%). The yellow-boxed area represents the area without data (corresponds to 6.25% of the total area). (c) The blue and red shaded lines represent the mean absolute error over training for the scenario with the full data and that with missing data, respectively. The results are obtained after training the neural network 100 times with different initial weights and biases. The line and shading represent the median and inter-quartile range, respectively. The inner figure shows the predicted elasticity distribution for the area without data based on the elasticity distribution on nearby regions.

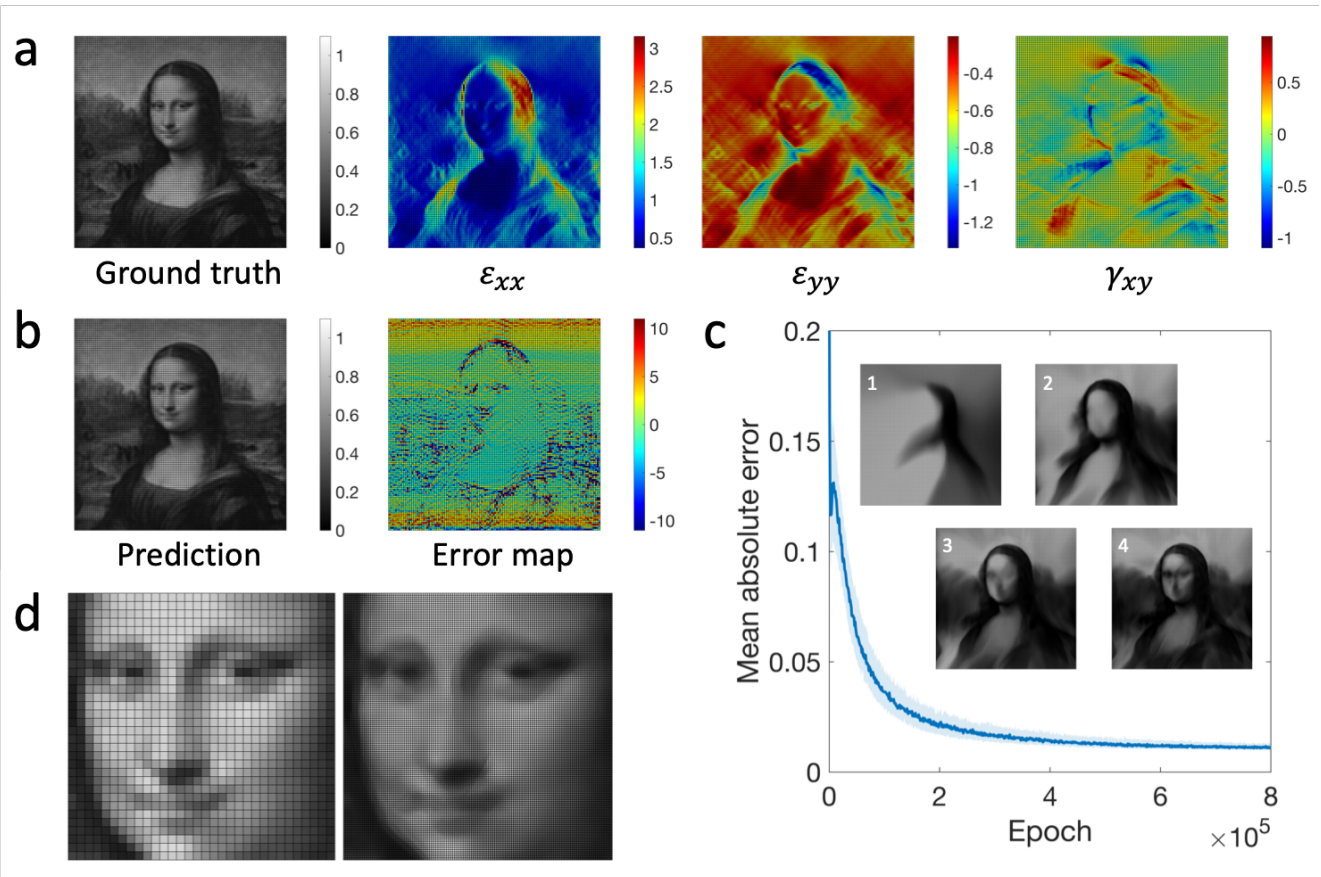

**Fig. 6| Super-resolution elasticity imaging.** (a) The elasticity and strain distributions of the Mona Lisa model. The unit in the elasticity image is MPa and that in the strain images is the percentage (%). (b) The predicted elasticity distribution and relative error map. The unit in the error maps is the percentage (%). (c) The blue shaded line represents the mean absolute error over training. The results are obtained after training the neural network 100 times with different initial weights and biases. The line and shading represent the median and inter-quartile range, respectively. The inner figures show the intermediate predictions after 280, 1,200, 2,100, 3,000 epochs, respectively. (d) A crop of the ground truth image (128-by-128) and that of the super-resolution image (512-by-512) generated by the neural network.

**Methods**

**The constitutive elasticity relation and conservation of linear momentum:** For a two-dimensional body, the relation between the strain and displacement with respect to the Cartesian axes is given by:

$$\boldsymbol{\varepsilon} = \begin{Bmatrix} \varepsilon_{xx} \\ \varepsilon_{yy} \\ \gamma_{xy} \end{Bmatrix} = \begin{Bmatrix} \dfrac{\partial u}{\partial x} \\ \dfrac{\partial v}{\partial y} \\ \dfrac{\partial u}{\partial y} + \dfrac{\partial v}{\partial x} \end{Bmatrix} \quad (7)$$

where $\boldsymbol{\varepsilon}$ is the strain vector, $u$ and $v$ are the horizontal and vertical component of the displacement, respectively. The constitutive elasticity relation for a linear elastic isotropic material in plane stress is given by:

$$\boldsymbol{\sigma} = \begin{Bmatrix} \sigma_{xx} \\ \sigma_{yy} \\ \tau_{xy} \end{Bmatrix} = \frac{E}{1-\nu^2} \begin{bmatrix} 1 & \nu & 0 \\ \nu & 1 & 0 \\ 0 & 0 & (1-\nu)/2 \end{bmatrix} \begin{Bmatrix} \varepsilon_{xx} \\ \varepsilon_{yy} \\ \gamma_{xy} \end{Bmatrix} \quad (8)$$

where $\boldsymbol{\sigma}$ is the stress vector, $E$ is the Young's modulus, and $\nu$ is the Poisson's ratio. Here, the Poisson's ratio is set to be 0.5 for incompressible materials. The conservation of linear momentum is typically written in a differential form:

$$\frac{\partial \sigma_{xx}}{\partial x} + \frac{\partial \tau_{yx}}{\partial y} = 0 \quad (9)$$

$$\frac{\partial \sigma_{yy}}{\partial y} + \frac{\partial \tau_{xy}}{\partial x} = 0 \quad (10)$$

These PDEs carry the derivatives of the stress field, which are the functions of the derivatives of the strain field and elasticity field. The derivatives of the strain field may be calculated from measured strains.

However, measured strains naturally contain noise and the calculation of the derivatives can amplify the noise significantly. The derivatives of the elasticity field cannot be calculated accurately when the derivatives of the strain field are inaccurate. To mitigate this potential problem, we rewrite the conservation of linear momentum in a finite difference form:

$$\frac{\sigma_{xx}(x+\Delta x, y) - \sigma_{xx}(x,y)}{\Delta x} + \frac{\tau_{yx}(x, y+\Delta y) - \tau_{yx}(x,y)}{\Delta y} = 0 \quad (11)$$

$$\frac{\sigma_{yy}(x, y+\Delta y) - \sigma_{yy}(x,y)}{\Delta y} + \frac{\tau_{xy}(x+\Delta x, y) - \tau_{xy}(x,y)}{\Delta x} = 0 \quad (12)$$

From (11) and (12), the conservation of linear momentum for a small cube with sides of length $h$ can be expressed in (1) and (2).

**Errors in predicted elasticity:** The error over training (Fig. 3c, Fig. 4c, Fig. 5c, and Fig. 6c) is quantified by the mean absolute error. The absolute error is defined as:

$$\text{error}_{\text{absolute}} = \sum_{i}^{m}\sum_{j}^{m} |E_{\text{pred}}(i,j) - E_{\text{truth}}(i,j)| \quad (13)$$

where $E_{\text{pred}}$ is the predicted elasticity, $E_{\text{truth}}$ is the true elasticity, and $m$ is the resolution of the elasticity image in both $x$-direction and $y$-direction. Here, $m$ is set to be 128. The absolute error, however, may not an ideal quantity to compare the errors between different models. A larger mean absolute error can be expected when the hidden elasticity is larger. Thus, the relative error is used to generate error maps (Fig. 3b, Fig. 4b, Fig. 5b, and Fig. 6b) and compare the accuracy between different models. The relative error (%) is defined as:

$$\text{error}_{\text{relative}} = 100 \times \sum_{i}^{m} \sum_{j}^{m} \left( E_{\text{pred}}(i,j) - E_{\text{truth}}(i,j) \right) / E_{\text{truth}}(i,j) \tag{14}$$

**Deep learning for elastography:** The neural network consists of 16 fully-connected hidden layers with 128 neurons per layer. The rectified linear unit is used as the activation function. The input of the neural network is a vector of two variables representing the position (*x* and *y*) of a material point and the output is the predicted elasticity. The stress is calculated by the encoded elastic constitutive relation from the measured strain and predicted elasticity jointly. A convolutional layer with the encoded kernels is used generate residual force maps. The Adam optimizer[28] is used to train the neural network to minimize the errors in the residual forces and physical constraint (either the traction on the boundary or the mean elasticity). The neural network is trained for 200,000 epochs for the inclusion model, sinusoidal model, and Cal model. For the Mona Lisa model, due to the extremely complex hidden elasticity, the neural network is trained for 800,000 epochs. The neural network has 248,193 parameters (weights and biases). As the weights and biases are randomly initialized before training, the predicted elasticity distribution cannot be exactly the same when training with different initial weights and biases. Thus, to better evaluate the performance of our new method, the predictions and errors reported in this study are the average values after training the neural network 100 times with different initial weights and biases. Our new method is deployed using TensorFlow[29] and training on the NVIDIA Tesla V100 and Titan V GPUs.

**Acknowledgements**

We acknowledge support from the NVIDIA GPU Seed Grant and Extreme Science and Engineering Discovery Environment (XSEDE), which is supported by National Science Foundation grant number ACI-1548562.